\documentstyle{mn}            

\title{Identification of a strong emission line at 2.8935$\mu$m in
 the spectrum of the Orion Nebula}

\author[L. B. Lucy]
       {L. B. Lucy \\
        Astrophysics Group, Blackett Laboratory, Imperial College of Science,
        Technology and Medicine, Prince Consort Road, London SW7 2BW\\
E-mail: l.lucy@ic.ac.uk}

\begin{document}

\maketitle

\begin{abstract}
A strong emission line at 2.8935$\mu$m discovered by Rubin et al. (2001) in
an ISO SWS02 spectrum of the Orion Nebula is identified as the  
$4p\; ^{3}\!P - 4s\; ^{3}\!S^{o}$ multiplet of O$\:${\sc i}. Line formation
is due to
de-excitation cascades following UV-pumping of high $^{3}\!S^{o}$ and   
$^{3}\!D^{o}$ terms and occurs in the O$\:${\sc i} zone immediately behind
the Hydrogen ionization front. This cascade mechanism also accounts for
permitted O$\:${\sc i} triplet lines in the Nebula's optical spectrum
(Grandi 1975).
An escape probability treatment of the O$\:${\sc i} cascades accounts for
the strength of the $\lambda$2.89$\mu$m line and suggests interesting
diagnostic possibilities for the optical lines.
\end{abstract}


\section{Introduction}

Rubin et al.(2001) have recently reported the detection of
an unidentified (uid)
strong emission line in an ISO SWS02 spectrum of the Orion Nebula. They
also report indications of the line's presence on two spectra from the ISO
archive taken at other locations in the Nebula and strong confirmation of the
line's reality in a UKIRT long-slit spectrum at the first location.
Assuming the line arises in the main ionization zone, they determine the
line's vacuum wavelength to be 2.89350$\mu$m. 

	Unidentified weak lines in astronomical spectra are of course common.
But this line is strong, being a factor of only 3.6 weaker than
the nearby H$\:${\sc i} 11-5 line at $\lambda$2.87$\mu$m. Given the efforts
of many
astronomers since the 1930's in studying the formation of emission
lines in photoionized 
nebulae, either as diagnostics or as coolants,
the discovery of a strong uid line at this late date is 
surprising and potentially of major significance.

	Rubin et al. (2001) report their own attempts to identify this line. 
They note satisfactory wavelength agreement for forbidden lines from two 
ionic species: Cr$\:${\sc iii} and Fe$\:${\sc v}, but argue against either
being the correct
identification. In this paper, an independent effort to identify this line
is reported,
which was started after seeing their IAU Abstract (Rubin et al. 2000).
The search for candidates was based on the Kurucz-Bell (1995)       
compilation of lines and $gf$-values.

\section{Required emissivity}
   
If, following Rubin et al (2001), we assume that the uid line
originates in the main ionization zone, then its emissivity can be
estimated from its strength relative to that of the nearby H line. 
Adopting $T_{e} = 9000$K and $N_{e} = 4000$cm$^{-3}$ as typical for the Orion
H$\:${\sc ii} region (Osterbrock, Tran \& Veilleux 1992) and setting
$N(H^{+}) = N_{e}$, we find
from Storey \& Hummer (1995) that the Case B
recombination emissivity of the 11-5 H$\:${\sc i} line at $\lambda$2.87$\mu$
is 
8.8 $\times$ 10$^{-21}$erg$\:$s$^{-1}$cm$^{-3}$. Accordingly, if the
unknown
emitting species has the same distribution as H$^{+}$, the
uid line's emissivity is   
\begin{equation}
 4 \pi j_{2.89\mu} \simeq 2.4  \times 10^{-21} erg\: s^{-1} cm^{-3} \;\;\; .
\end{equation}
Any proposed identification for which the emission
originates within the H$\:${\sc ii} region must be capable of emitting at
this rate.

\section{Rejected candidates}

	In this section, candidates found on the basis of wavelength 
coincidence are reported and reasons given for their rejection. Because
this work partly duplicates that of Rubin et al. (2001), the discussion is
abbreviated.

	An excellent candidate on the basis of wavelength is a forbidden
line of Chromium at 2.89349$\mu$m. This was discarded immediately because
of the low cosmic abundance of this element. Rubin et al. were more
persistent but eventually also concluded against this identification,
with the stronger argument that three other [Cr$\:${\sc iii}] lines in the
IR are not seen in their spectra.

	Another similarly acceptable candidate is a forbidden line of Fe$\:${\sc v}
at $\lambda$2.89343$\mu$m. Rubin et al. reject this candidate since the ionization
potential of Fe$\:${\sc iv} (54.8eV) slightly exceeds that of He$\:${\sc ii}
(54.4eV), implying a negligible rate of creation of Fe$\:${\sc v}
ions by photoionization. But, given that Rubin et al. (2000) had evidently
failed to find an identification from among the ions expected to be 
abundant in the H$\:${\sc ii} region, the Fe$\:${\sc v} possibility was
pursued further.

	Statistical equilibrium calculations
to predict emissivities for the rich spectrum of
[Fe$\:${\sc v}] lines were carried out using the
A-values of Garstang (1957) and the collision strengths of Berrington (2001).
For $T_{e}= 9000$K and $N_{e} = 4000$cm$^{-3}$, the emissivity per
Fe$\:${\sc v} ion of the transition
$3d^{4} \; ^{3}\!G_{5} -3d^{4} \; ^{3}\!F2_{4}$ at $\lambda$2.89343$\mu$m
is 0.93$\times$10$^{-20}$ erg$\:$s$^{-1}$ cm$^{-3}$. Accordingly, to
achieve the target emissivity given in equation (1), the number density of
Fe$\:${\sc v} ions must be $\simeq$ 0.26cm$^{-3}$. But at
$N_{e} = 4000$cm$^{-3}$ and with an Fe abundance $n(Fe)/n(H) = -4.49$dex
(Seaton et al. 1994), the number density of Fe {\em atoms} is only
$\simeq$ 0.13cm$^{-3}$. If further evidence against this identification
were needed, these calculations reveal that there are several other 
[Fe$\:${\sc v}] lines in the IR with greater emissivities, including two
from the same upper level as this rejected candidate.

\section{Proposed candidate}

The suggested identification is the O$\:${\sc i} multiplet
$4p\; ^{3}\!P - 4s\; ^{3}\!S^{o}$. The permitted components of this multiplet are at
vacuum wavelengths 2.89330, 2.89352 and 2.89359$\mu$m with gf-values
of 0.50, 2.51 and 1.50, respectively.

	Two mechanisms of line formation will be considered in this section
in an attempt to justify this identification. The aim is to
demonstrate that emission in this transition 
can plausibly account for the maximum observed line brightness. No attempt
is made to give a definitive treatment of line formation.  
 
\subsection{Atomic data}
 
The atomic model is restricted to the O$\:${\sc i} triplet system.
Specifically, all terms in the $^{3}\!S^{o},\; ^{3}\!P,\; ^{3}\!D^{o}$ and
$^{3}\!F$
series are included up to principal quantum numbers $n =$ 11, 10, 11 and 10,
respectively. Energy levels are the experimental values from the NIST Atomic
Spectra Database if available. For higher $n$, we take 
\begin{equation}
 E_{n} = E_{\infty} - \frac{R_{O}}{(n-a)^{2}}     \;\;\; ,
\end{equation}
where $E_{\infty}$ is the ionization potential, $R_{O}$ is the Rydberg
constant for Oxygen, and $a$ is a constant for a given series obtained by
fitting this formula to
the highest level reported in the NIST database. The resulting atomic model
has 84 levels. Note that sublevels belonging to the same term are {\em not}
consolidated into a single level.

	Oscillator strengths for all the permitted multiplets in the above
O$\:${\sc i} model were computed by Butler \& Zeippen
(1991) as part of the Opacity Project and are available in the TOPbase
archive. Values for the individual
transitions were derived assuming LS-coupling.  

	Radiative recombination coefficients for the individual triplet terms 
for a wide range of electron temperatures have been computed by Nahar (1999).
Values for the sublevels are derived assuming proportionality to their
statistical weights.

	Grotrian diagrams for the O$\:${\sc i} triplet system have been
published by Grandi (1975) and Przybilla et al. (2000).

\subsection{Recombination cascade}

By analogy with the H emission-line spectra of H$\:${\sc ii} regions and
planetary
nebulae, the most obvious emission mechanism for this O$\:${\sc i}
multiplet is via a
radiative cascade following recombinations of O$^{+}$ ions to high triplet
levels.

	Because the ionization potentials of O$\:${\sc i} (13.618 eV) and
H$\:${\sc i} 
(13.598 eV) are essentially identical, the emitting volumes for
O$\:${\sc i} and H$\:${\sc i}
recombination lines are also identical, namely the H$\:${\sc ii} region.
Accordingly,
on the assumption that it is a recombination line, the viability of this
identification can be tested by comparing the O$\:${\sc i} multiplet's
emissivity with the estimate given in equation (1).

	The emissivity is calculated at
$N_{e} = 4000$cm$^{-3}$ and
$T_{e} = 10^{4}$K, this being a temperature at which Nahar (1999) tabulates
recombination coefficients. The Oxygen abundance 
is taken to be $n(O)/n(H) = -3.38$dex, the B star value adopted by
Savage \& Sembach (1996). 

       The calculation proceeds as follows: the rate
of recombinations to level $i$ per unit volume is
$\alpha_{i} n(O^{+}) n_{e}$, 
where $\alpha_{i}$ is the recombination coefficient for level $i$. These
recombinations create a cascade down to the ground term. Accordingly, let
$\dot {R}_{i}$
denote the rate per unit volume at which level $i$ is populated by
recombinations directly to
$i$ {\em and} by cascades from higher levels. Then, since the fraction 
\begin{equation}
   p_{ij} = A_{ij} / \sum_{\ell} A_{i \ell}     \;\;\; 
\end{equation}
of the population of level $i$ decays to level $j<i$, where $A_{ij}$ is
the Einstein A-value for the transition $i \rightarrow j$, the emissivity
of this transition is  
\begin{equation}
   4 \pi j_{ij} =  p_{ij} \dot {R}_{i} h \nu_{ij}     \;\;\; .
\end{equation}
Moreover, these $i \rightarrow j$ decays make the contribution  
$p_{ij} \dot {R}_{i}$ to the quantity $\dot {R}_{j}$.

	The above calculation is started by setting 
$\dot {R}_{u} = \alpha_{u} n(O^{+}) n_{e} $, where $u$ denotes the highest
level in the atomic model, and then proceeds downwards level by level
to the ground term, resulting in emissivities for all permitted lines
in the recombination cascade.

	For the three lines contributing to the
$4p\; ^{3}\!P - 4s\; ^{3}\!S^{o}$ multiplet, the total emissivity is
5.6 $\times$ 10$^{-24}$ erg$\:$ s$^{-1}$ cm$^{-3}$, a mere 0.23 per cent
of the
required value. But this is a case A calculation in
which photons emitted in decays to the ground term escape.
For case B, which is obtained by repeating the above calculation with
$A_{i \ell} = 0$ for $\ell = 1-3$, the emissivity increases to
2.3  $\times$ 10$^{-23}$ erg$\:$ s$^{-1}$ cm$^{-3}$, still only 0.96 per
cent of the required value. 
 
\subsection{Pumped cascades}

If emission in the $4p\; ^{3}\!P - 4s\; ^{3}\!S^{o}$  O$\:${\sc i}
multiplet were possible only following recombinations of O$^{+}$ ions, then
this candidate would now be decisively ruled out. But
since the ground term
of O$\:${\sc i} is also a triplet, $\lambda 2.89 \mu$m emission can also
occur via UV pumping of high triplet
levels. Accordingly, the suggested
emission region is the neutral O zone
which, because of the essentially identical ionization potentials noted
earlier, is coextensive with the H$\:${\sc i} region - see Fig. 1. In this
region,
the far UV but non-ionizing radiation - ie. $\lambda > 911$ \AA\ -  of
$\theta^{1}$ Ori C penetrates and excites O$\:${\sc i} atoms
from the
three levels
of the ground term to high $^{3}\!S^{o}$ and $^{3}\!D^{o}$ terms, from whence
radiative decays give rise to emission in the
$\lambda$ 2.89 $\mu$m multiplet. Exactly the same mechanism was invoked by
Grandi (1975) to explain the permitted O$\:${\sc i} triplet
lines in optical spectra of the Orion Nebula.

\begin{figure}
\vspace{8.2cm}
\includegraphics{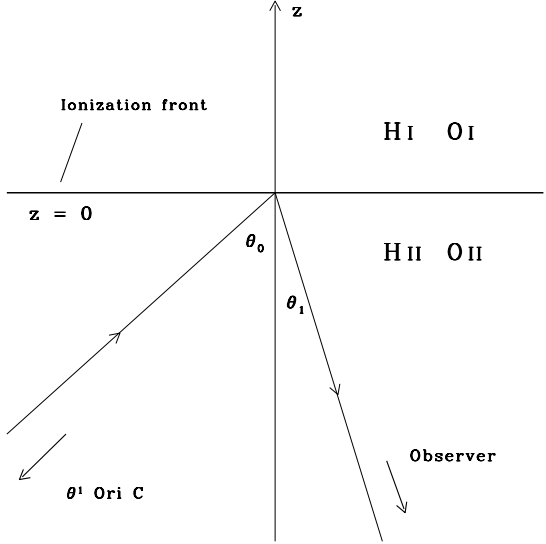}
\caption{Geometrical configuration assumed for the radiative transfer
calculation. Radiation from $\theta^{1}$ Ori C is incident
at angle $\theta_{0}$ on a semi-infinite uniform layer
of neutral
Hydrogen and Oxygen. Line emission created in the pumped cascades 
propagates towards the Observer at angle $\theta_{1}$.}

\end{figure}

	For this line-formation mechanism, the H$\:${\sc ii} region is not
the source of emission; and so 
the emissivity estimate in equation (1) is 
no longer the relevant target. Instead, the proposal must be
tested by predicting the line's surface brightness.
In particular, a viable identification must account for
the maximum brightess measured by Rubin et al. (2001),    
\begin{equation}
  f_{2.89 \mu} = 0.12\: bu   \;\;\; ,
\end{equation}
where a brightess unit (bu) of
10$^{-13}$ erg$\:$ s$^{-1}$ cm$^{-2}$ arcsec$^{-2}$ is adopted.

	In the remainder of
this section, a crude theory is developed with the deliberate aim of   
overestimating the line's brightness. If this overestimate were to fall short
of this maximum, a more detailed treatment would not be necessary.

	In computing the line's brightness, we assume that the UV
pumping and subsequent de-excitation cascades occur in a 
semi-infinite uniform slab of neutral O atoms whose surface $z=0$ is
coincident
with the Hydrogen ionization front. The incident pumping radiation from
$\theta^{1}$ Ori C has direction cosine $\mu_{0} = cos \: \theta_{0}$, and
the observer receives radiation emerging at $\mu_{1} = cos \: \theta_{1}$.  

	Let us consider the pumping line at frequency $\nu_{k}$
corresponding to one of the permitted transitions $\ell \rightarrow u$ for
$\ell = 1-3$ and $u \geq 4$. On the
assumption of zero attenuation in the H$\:${\sc ii} region by either dust
or neutral O 
atoms, the intensity at $z=0$ of the radiation from
$\theta^{1}$ Ori C is $I_{k}=F(\nu_{k})$, where $\pi F(\nu_{k})$ is the
emergent flux at $\nu_{k}$ from the star's atmosphere. Accordingly, the
rate per unit area at which energy in the frequency interval
$(\Delta \nu)_{k}$ centred on $\nu_{k}$  
crosses into the O$\:${\sc i} zone is
\begin{equation}
   \dot {E}_{k} = I_{k}\: \mu_{0}\: (\Delta \nu)_{k} (\Delta \omega)^{*}  \;\;\; .
\end{equation}
Here $ (\Delta \omega)^{*} = \pi (R/d)^2$ is the solid angle subtended by 
$\theta^{1}$ Ori C from the point considered at the ionization front.  
Note that the contribution of diffuse radiation from the  H$\:${\sc ii}
region is neglected.

	If we now assume that all the energy in the interval 
$(\Delta \nu)_{k}$
is absorbed in the slab by neutral O atoms, then $\dot {E}_{k}/h \nu_{k}$
is the rate 
at which the $\nu_{k}$ line photons excite level $u$ per unit area. Let us
further suppose that each of these pumping events leads on average to the
emission
of $\epsilon^{k}_{ij}$ ($<$1) cascade photons of frequency $\nu_{ij}$. It then
follows that
\begin{equation}
   4 \pi \int_{0}^{\infty} \!\! j_{\nu_{ij}} dz = \epsilon^{k}_{ij}\: \frac{\nu_{ij}}{\nu_{k}}\:
\dot {E}_{k}  \;\;\; ,
\end{equation}
where $4 \pi j_{\nu_{ij}}$ is the emissivity of the line $\nu_{ij}$.
 
	Because the cascade photons of interest are in the IR and are emitted
by subordinate lines, we can neglect attenuation as they propagate to the
surface of the slab. The transfer equation to be solved is therefore simply
\begin{equation}
	\mu \frac{d I_{\nu}}{d z} =  j_{\nu_{ij}}\: \phi_{\nu}  \;\;\; , 
\end{equation}
where $\phi_{\nu}$ is the cascade line's normalized emission profile.
For a uniform O$\:${\sc i} slab, this profile is independent of $z$, and so
the emergent intensity is 
\begin{equation}
     I_{\nu}(\mu) = \frac{1}{\mu}\: \phi_{\nu} \int_{0}^{\infty} \!\! j_{\nu_{ij}}\: dz  \;\;\; . 
\end{equation}

	Now if $\vartheta \times \vartheta $ is the element of sky used to
define surface brightness, then, neglecting extinction
between the ionization front and the observer, the line brightness at 
frequency $\nu$ is  $I_{\nu}(\mu_{1})\: \vartheta^{2}$. Accordingly, if
we integrate over the line profile, the contribution to the brightness of
the cascade line $i \rightarrow j$ from the $k$th pumping transition is  
\begin{equation}
      f^{k}_{ij}= \frac{\vartheta^{2}}{\mu_{1}} \int_{0}^{\infty} j_{\nu_{ij}} dz  \;\;\; .  
\end{equation}

	Let us now suppose that isotropic macroturbulence with
characteristic velocity
$v_{D}$ is the cause of kinematic broadening in the  O$\:${\sc i} slab and
write 
$(\Delta \nu)_{k} = w_{k} \Delta \nu_{D}$, so that $w_{k}$ is the
dimensionless
equivalent bandwidth for absorption by the $k$th pumping line. Then,
using equation (7) to eliminate the emissivity integral and substituting
$\dot {E}_{k}$ from equation (6), we obtain
\begin{equation}
      f^{k}_{ij}= \frac{1}{4} \vartheta^{2} \times \frac{\mu_{0}}{\mu_{1}}
(\frac{R}{d})^{2} \times \epsilon^{k}_{ij}  w_{k} \times 
\frac{\nu_{ij}}{\nu_{k}} F(\nu_{k}) \Delta \nu_{D}
    \;\;\; . 
\end{equation}

	Equation (11) gives the contribution to the strength of the line
$i \rightarrow j$ from the
cascade driven by the $k$th pumping transition. For the
$\lambda 2.89 \mu$m multiplet,
these pumping
transitions comprise all the permitted transitions
connecting the three states of the ground term
$2p^{4} \: ^{3}\!P$ to the states belonging to the terms $ns\: ^{3}\!S^{o}$
with
$n \geq5$ and to the terms $nd ^{3}D^{o}$ with $n \geq4$.
Summation of equation (11)
over these pumping transitions gives the required line strength.

	 In the absence of
predictions for the equivalent bandwidths and for the efficiencies with which
cascade
photons are emitted, we make the optimistic estimates that, for all $k$,
$w_{k}=5$ and $\epsilon^{k}_{ij}=1$. In addition, we assume blackbody emission
by $\theta^{1}$ Ori C with $T_{eff}=40000K$, and take $R=11.1R_{\sun}$ (Lucy 1995).

	Because Rubin et al. (2001) observed maximum line brightness at 
position 1SW, a line-of-sight that passes $\theta^{1}$ Ori C with a small 
impact parameter of $ \simeq 0.08 pc $, we set
$d = 0.2 pc$, which is Wen \& O'Dell's (1995) estimate of this star's
distance above the ionization front. In addition, since the line-of-sight
with zero impact parameter has $\mu_{0} = \mu_{1}$, we take the
orientation factor $\mu_{0}/\mu_{1} = 1$.

	Finally, the macroturbulent velocity $v_{D}$ must be estimated.
From their UKIRT spectrum, Rubin et al. (2001) measure the resolved FWHM of
the $\lambda 2.89 \mu$m line to be 24 km s$^{-1}$. If the line were single,
this would give $v_{D} =$ 14.4 km s$^{-1}$. But allowing for the three 
unresolved components with relative strengths from Section 5, we obtain
$v_{D} =$ 12.7 km s$^{-1}$, a value that will be used throughout this
investigation. Of course, the actual broadening may be dominated by a 
differentially expanding flow, with a much smaller turbulent component.
Evidence of non-thermal broadening of other lines in the spectrum of 
the Orion Nebula are summarized by O'Dell (2001b).

	With these choices of the parameters, the predicted line brightness
is
\begin{equation}
 f_{2.89 \mu} = 1.48 \: bu \;\;\; ,
\end{equation}
which is a factor 12.3 greater than the maximum of
0.12 bu
observed by Rubin et al (2001). By thus comfortably exceeding observed
line strengths, the proposed identification survives this initial
test. In fact, in the absence of alternatives, this is already a
powerful argument in favour of the O$\:${\sc i} multiplet.

	Note that, according to this crude theory, the brightness is
independent of the Oxygen abundance. This arises because of the assumption
that all photons within a fixed bandwidth are absorbed by the pumping lines.

\section{Improved theory}

	The simple calculation of Sect. 4.3 is consistent with the
identification of
the $\lambda$2.89$\mu$m line as the O$\:${\sc i} multiplet 
$4p\; ^{3}\!P - 4s\; ^{3}\!S^{o}$. Now an escape probability
treatment is developed to see if this identification 
remains viable when the optimistic estimates of
$w_{k}=5$ and $\epsilon^{k}_{ij}=1$ are replaced by actual
calculations. In
addition, if the identification is confirmed, a more realistic theory will be
needed to exploit the measured strengths of this and other
O$\:${\sc i} lines
formed by the cascade mechanism.

\subsection{Incident intensities}
  
	In Sect. 4.3, the incident intensity
$I_{k} = B_{\nu_{k}}(T_{eff})$,
with T$_{eff}$ = 40000K. But the pumping lines are in the wavelength
range 918-979 \AA\ where line blocking by the Lyman series
and by metal lines is expected to be strong. Accordingly, the first
improvement is to replace the black body approximation by the
Kurucz (1979) LTE line-blanketed atmosphere with
T$_{eff}$ = 40000K, log $g$ = 4.5 and solar metal abundances. This model
was previously used to investigate the fluorescent excitation of
[Ni$\:${\sc ii}] and [Fe$\:${\sc ii}] lines in the Orion Nebula
(Lucy 1995).   

	As a result of this change, the incident intensities for lines with 
$\lambda <$ 930 \AA\ 
are reduced by up to 25 percent while those with longer
wavelengths are increased by up to 60 percent. Yet more accurate  
would be a NLTE model incoporating spherical extension 
and the stellar wind (Gabler et al. 1989).

\subsection{Equivalent bandwidths}

The rate per unit area at which the $k$th pumping line absorbs energy from
the incident beam  is
\begin{equation}
  \dot{E}_{k} = \int_{0}^{\infty} \!\! dz \;\; 4\pi \!\int \ell_{\nu} J_{\nu} d \nu                                     \;\;\; ,
\end{equation}
where  $\ell_{\nu}$ is the line's absorption coefficient per unit volume
and  $J_{\nu}$ is the beam's mean intensity at
depth $z$.

	On the assumption that the pumping is confined to a layer whose
thickness is small compared to the distance $d$ of $\theta^{1}$ Ori C,
$J_{\nu}= I_{\nu} \times  (\Delta \omega)^{*}/4 \pi$, where $I_{\nu}$,
the beam's specific intensity, is given by
\begin{equation}
   I_{\nu}= I_{k}\: exp[-(k_{\nu}+\ell_{\nu}+m_{\nu})\: \frac{z}{\mu_{0}}]
\end{equation}
Here $k_{\nu}$ is the absorption coefficient of unit slab volume due to
interstellar grains and $m_{\nu}$ is the summed absorption coefficients of any
other pumping lines that overlap the line considered, and which therefore
compete for photons.

	If we substitute for $J_{\nu}$ in equation (13) and integrate,
we recover equation (6) except that $(\Delta \nu)_{k}$ is now replaced by
$w_{k} \Delta \nu_{D}$, with equivalent bandwidth $w_{k}$ given by
\begin{equation}
 w_{k} = \frac{1}{\Delta \nu_{D}} \int 
\frac{\ell_{\nu}}{k_{\nu}+\ell_{\nu}+m_{\nu}} \: d \nu       \;\;\; .
\end{equation}

\subsubsection{Dust absorption}

The extinction coefficient for a typical mixture of interstellar
grains can be derived from standard fits to observational data as follows: a 
line of length $s$ in the O$\:${\sc i} slab corresponds to optical depth
$\tau_{\lambda} = k_{\lambda}^{ext} s = 0.4 \ell n 10 \times  A(\lambda)$,
where
$A(\lambda)$ is the 
extinction at $\lambda$ in magnitudes. This can be expressed in terms of the
colour excess $E(B-V)$ by writing 
$A(\lambda)= A(\lambda)/A(V) \times R_{V} \times E(B-V)$, where
$R_{V}=A(V)/E(B-V)$ is the ratio of visual to selective extinction. But
the colour excess is found to be proportional to the H column density. Thus,
we also have $N(H) = n(H)s = \gamma E(B-V)$, with 
$\gamma \simeq 5.8 \times 10^{21}$ atoms cm$^{-2}$ mag$^{-1}$ (Bohlin, Savage
\& Drake 1978). Eliminating $s$ from these formulae, we find that
\begin{equation}
 k_{\lambda}^{ext}= 0.4 \ell n 10 \times \frac{A(\lambda)}{A(V)} \times R_{V}
            \times \frac{n(H)}{\gamma}      \;\;\; ,
\end{equation}
This is evaluated using the equation  $A(\lambda)/A(V) = a(x)+b(x)/R_{V}$,
where $x$ is the reciprocal wavelength in $\mu m^{-1}$ 
and the functions $a$ and $b$ are the fits to observational data given
by Cardelli, Clayton \& Mathis (1989).
Note that since the pumping lines are such that
$10.2 < x < 10.9$, this application implies a modest extrapolation
beyond the observed extinction data. 
The ratio $R_{V}$ is set = 3.1, the
standard value for the diffuse interstellar medium (eg, Cardelli et al. 1989).

	Equation (16) gives the sum of the dust absorption and the scattering 
coefficients. If
$\varpi_{\lambda}$ is the albedo at $\lambda$, the absorption
coefficient is
\begin{equation}
 k_{\lambda}= (1-\varpi_{\lambda}) k_{\lambda}^{ext}     \;\;\; .
\end{equation}
In this investigation, we set $\varpi_{\lambda} = 0.4$ for all the 
far-UV pumping lines. This value is suggested by Fig. 8 in Gordon et al.
(1994), which summarizes the work of these and earlier workers.

\subsubsection{Line absorption}
 
On the assumption that isotropic macroturbulence with characteristic
velocity
$v_{D}$ is the dominant broadening mechanism, the normalized line absorption
profile is
\begin{equation}
  \phi_{\nu} = \frac{1}{\sqrt{\pi}}\: \frac{1}{\Delta \nu_{D}} \: 
               exp[-(\frac{\nu-\nu_{k}}{\Delta \nu_{D}})^{2}]    \;\;\; ,
\end{equation}
where  
$\Delta \nu_{D}/\nu_{k} = v_{D}/c$. The resulting line absorption coefficient
per unit volume is
\begin{equation}
  \ell_{\nu} = \frac{\pi e^{2}}{m_{e} c^{2}}\: f_{\ell u}  
          \times \frac{n_{\ell}}{n(O)} \times n(O) \times \phi_{\nu}  \;\;\; .
\end{equation}
Here the index $\ell = 1-3$ refers to the three levels of the ground term, 
and the index $u$ refers to the pumped levels belonging to the series 
$ns \; ^{3}\!S^{o}$ and $nd \; ^{3}\!D^{o}$.

\subsubsection{Computed bandwidths}

	Before calculating the quantities $w_{k}$ from equation (15), line
overlaps must be identified. No overlaps occur for the multiplets
$ 2 p^{4}\; ^{3}\!P \rightarrow ns\; ^{3}\!S^{o}$ because the lower levels are
well
separated and the upper levels are single. But for the
multiplets $ 2 p^{4}\; ^{3}\!P \rightarrow nd \; ^{3}\!D^{o}$ overlaps do
occur
because of the
small fine structure splittings of the upper terms. Specifically, the three
components
$2 p^{4}\; ^{3}\!P_{2} \rightarrow nd \; ^{3}\!D^{o}_{1,2,3}$ compete for
photons, as do the two components   
$2 p^{4}\; ^{3}\!P_{1} \rightarrow nd \; ^{3}\!D^{o}_{1,2}$.

	In contrast to the calculation of Sect. 4.3, which was independent 
of the Oxygen abundance,
this improved theory depends on $n(O)/n(H)$ through the ratios
$\ell_{\nu}/k_{\nu}$ and $m_{\nu}/k_{\nu}$. As in Section 4.2, we take
$n(O)/n(H)=-3.38$ dex. 

	In addition, the populations
of the levels comprising the ground term must be specified. Assuming
proportionality to statistical weights, we set
$n_{\ell}/n(O) = (2J+1)/9$. In fact, these values are in
fair accord with the detailed calculations of Pequignot (1990) for the
expected
conditions immediately behind the ionization front - i.e., 
n(H) $\ga 10^{5}$ cm$^{-3}$, T $\ga$ 5000 K, n$_{e}$/n(H) $\la$ 0.01.

	The equivalent bandwidths $w_{k}$ obtained from equation (15)
for all relevant
pumping lines are plotted in Fig. 2. These widths are all less than the
optimistic estimate of 5 used in Sect.4.3. Nevertheless, a substantial
number of lines have $ w_{k} \ga 2$, thus indicating effective pumping.

\begin{figure}
\vspace{8.2cm}
\includegraphics{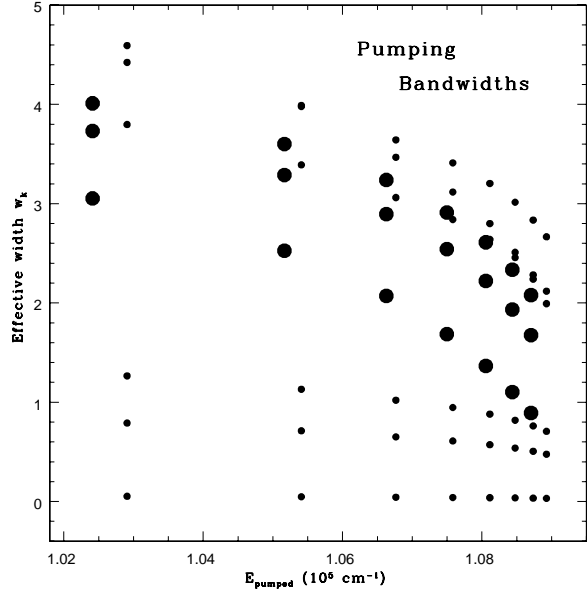}
\caption{Equivalent bandwidths $w_{k}$ from
equation (15) plotted against the excitation energy of the pumped level.
Results are shown for all the pumping transtions contributing to emission in
the O$\:${\sc i} $\lambda$2.89 $\mu$m
multiplet. Large filled circles identify transitions whose upper levels
are $n \; ^{3}\!S_{1}^{o}$ for $n = 5-11$. Small filled circles identify
transitions whose upper levels are  $n \; ^{3}\!D^{o}_{1,2,3}$ for
$n=4-11$.}
\end{figure}

	The general trends shown in Fig. 2 are readily understood. Each
$^{3}\!S^{o}$ term contributes three lines and their $w_{k}$ values fall with
increasing $n$ due to declining $gf$-values, which imply greater
losses to dust absorption. For the six lines from each 
$^{3}\!D^{o}$ term, there is the additional effect of line overlap. As a
result, overlapped lines with small $gf$-values are starved
of photons and thus make little contribution to the pumping. This effect
explains why, for each $n$, three of the $n \; ^{3}\!D^{o}$ components have 
$w_{k} \la 1.3$.

\subsection{Escape probabilities}

In cascade calculations, one of two extreme cases is usually 
considered: case A, in which photons emitted in decays to the ground term
escape the nebula; or case B, in which all such photons are
re-absorbed by the emitting transitions. In contrast, here we
use escape probabilities to treat intermediate circumstances. 

	For subordinate transitions - i.e. $ u \rightarrow \ell$ for
$ \ell > 3$ , the probability of re-absorption is negligible, and so 
we take their escape probabilities $ \beta_{u \ell} =1$. But for decays
to the ground term $2p^{4}\;^{3}\!P$, the possibility of re-absorption needs
to be 
considered. If a photon of frequency $\nu$ is emitted in the decay to level
$2p^{4} \; ^{3}\!P_{J}$, the probability of it being absorbed by a dust grain
is $k_{\nu}/(k_{\nu}+\ell_{\nu}+m_{\nu})$, where the notation is that of
Section 5.2. Thus, by averaging over the line's emission profile, we find
that the escape probability is
\begin{equation}
 \beta_{u \ell} = \int 
\frac{k_{\nu}}{k_{\nu}+\ell_{\nu}+m_{\nu}}\: \phi_{\nu} \: d \nu  
\end{equation}
for $ \ell = 1-3 $ and $u \geq 4$.

	If a photon emitted in a decay to the ground term is not absorbed
by a dust grain, it is here assumed to be re-absorbed by the emitting
transition.
But this neglects two effects. First, there is the additional 
possibility of
escaping by crossing back into the H$\:${\sc ii} region - i.e. $z < 0$ in
Fig. 1. Second, there is the possibility - for decays from the $^{3}\!D^{o}$
terms - of being absorbed by an overlapping component, thus 
transferring excitation to a different sublevel of the same $^{3}\!D^{o}$
term.

	The first possibility is neglected to avoid a $z-$dependent 
cascade calculation, a complexity not warranted at this point. The
second possibility is neglected since excitation is not lost, merely
transferred. As such, this assumption should have little effect on
predicted line strengths. (In fact, a code incorporating
this interlocking effect has been written, and this statement is confirmed.) 

	In computing escape probabilities with equation (20), 
$k_{\nu}$ is calculated exactly as in Section 5.2.1.
But
the calculation of the line absorption coefficients $\ell_{\nu}$ and
$m_{\nu}$ and the emission profile $\phi_{\nu}$ departs slightly from Section
5.2.2 in that the macroturbulent velocity $v_{D}$ is replaced by
$\sqrt{2kT/m_{O}}$, the
thermal velocity of the Oxygen atoms. This allows for
the possibility that the length scale on which photon recapture occurs is
less than that of the bulk motions that determine the broadening of
the $\lambda 2.89 \mu$m line. The temperature immediately behind the
ionization front is taken to be $T = 10000$K
(Esteban, Peimbert \& Torres-Peimbert 1999).   

	The resulting escape probabilities are plotted in Fig.3. As for
Fig.2, the general trends are readily understood as being due to decreasing
$gf$-values as one ascends the two series. But one point of difference
is that the $^{3}\!D^{o}$ terms provide three points on Fig.3 as against six
on Fig.2. This is because the quantity $\ell_{\nu}+m_{\nu}$ in equation (20) 
is identical for the three overlapping lines
$nd \; ^{3}\!D^{o}_{1,2,3} \rightarrow 2 p^{4}\; ^{3}\!P_{2}$, which therefore
have identical escape probabilities. Similarly, the two overlapping lines 
$nd \; ^{3}\!D^{o}_{1,2} \rightarrow 2 p^{4}\; ^{3}\!P_{1}$ also
have identical escape probabilities.

\begin{figure}
\vspace{8.2cm}
\includegraphics{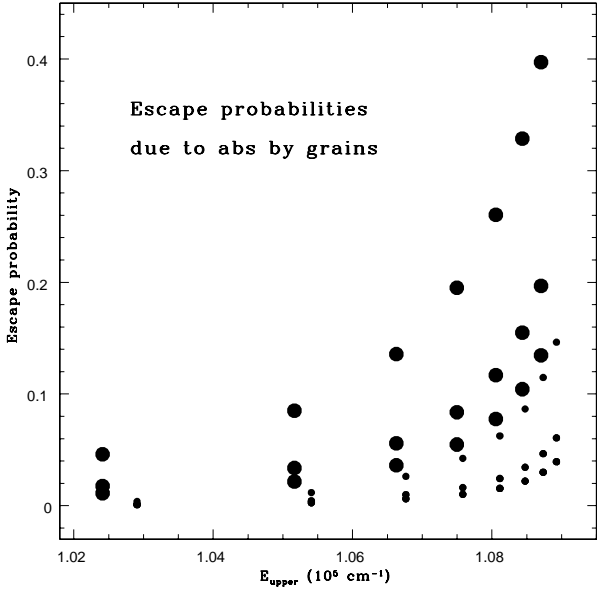}
\caption{Escape probabilities $\beta_{u \ell}$ from equation (20)
plotted against the excitation energy of the emitting level.
Large filled circles identify transitions whose upper levels
are $n \; ^{3}\!S_{1}^{o}$ for $n = 5-11$. Small filled circles identify
transitions whose upper levels are  $n \; ^{3}\!D^{o}_{1,2,3}$ for
$n=4-11$. Line absorption coefficients are computed on the assumption of
Doppler broadening by thermal motions at $T = 10^{4}K$.}
\end{figure}

\subsection{Efficiencies}

With escape probabilities determined, the efficiencies
$\epsilon^{k}_{ij}$ can now be derived from the cascade
driven by the $k$th pumping transition.
The calculation of the pumped cascades follows closely that for
recombinations described in Section 4.2.

	The photon absorption rate per unit slab area in the
$k$th
pumping line is
$\dot{E}_{k}/h \nu_{k}$. The quantity $\dot{E}_{k}$ is derived
from
equation (6) with $\mu_{0} =1$ and $(\Delta \nu)_{k} = w_{k} \Delta \nu_{D}$,
with equivalent bandwidths from Section 5.2.

	If the $k$th pumping line corresponds to the transition  
$\ell \rightarrow u$, then the $k$th cascade starts at level $u$ with
$\dot{R}_{u} = \dot{E}_{k}/h \nu_{k}$.
From a level $i \leq u$, the fraction of
decays that go to level $j$ is 
\begin{equation}
   p_{ij} = A_{ij} \beta_{ij} / \sum_{\ell} A_{i \ell} \beta_{i \ell} \;\;\; ,
\end{equation}
where the escape probabilities are from Section 5.3; and these decays make
the contribution $p_{ij} \dot{R}_{i}$ to $\dot{R}_{j}$. As in Section 4.2,
this calculation proceeds downwards level by level until the ground term
is reached.  

	When the cascade terminates, photon emission rates $p_{ij}\dot{R}_{ij}$ are
available
for all permitted cascade transitions $i \rightarrow j$. But these rates were
achieved by absorbing pumping photons at the rate
$\dot{R}_{k}=\dot{E}_{k}/h\nu_{k}$. Accordingly,
the efficiency with which the $k$th pumping line creates $i \rightarrow j$
photons is
\begin{equation}
\epsilon^{k}_{ij}= p_{ij} \dot{R}_{i} /  \dot{R}_{k}  \;\;\; .
\end{equation}
Note that because of the cascade's linearity the efficiencies
$\epsilon^{k}_{ij}$
are independent of $\mu_{0}$.

\begin{figure}
\vspace{8.2cm}
\includegraphics{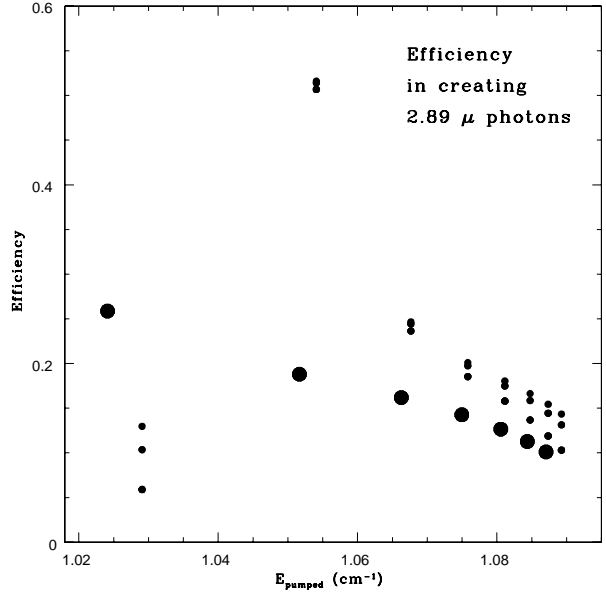}
\caption{Efficiencies $\epsilon^{k}_{2.89 \mu}$ from equation (22)
plotted against the excitation energy of the pumped level.
Large filled circles identify transitions whose upper levels
are $n \; ^{3}\!S_{1}^{o}$ for $n = 5-11$. Small filled circles identify
transitions whose upper levels are  $n \; ^{3}\!D^{o}_{1,2,3}$ for
$n=4-11$.}
\end{figure}

	The efficiencies with which the numerous pumping lines create
$\lambda$2.89 $\mu$m photons are plotted in Fig.4. The highest efficiencies of
$\simeq 0.51$ are found for the six components of the multiplet 
$2p^{4} \; ^{3}\!P \rightarrow 5d \; ^{3}\!D^{o}$. But most are in the
range 0.1-0.2, a substantial reduction from the optimistic estimate of
$\epsilon^{k}_{2.89 \mu}=1$ used in Section 4.3.

	Surprisingly low efficiencies are found for the components of the
multiplet $2p^{4} \; ^{3}\!P \rightarrow 4d \; ^{3}\!D^{o}$, with values in
the range 0.06-0.13. Inspection of the branching ratios for decay channels
from the $4\; ^{3}\!D^{o}$ levels shows that, despite the small escape
probabilities for decays to the ground term - see Fig.3 - these still
dominate, accounting for 55-80 per cent of the total decay rates from these
levels.

	The efficiencies plotted in Fig.3 are not an immediately reliable
guide to the relative importance of the numerous pumping transtions since
a line of high efficiency might have a low equivalent bandwidth. Accordingly,
the contribution of each pumped term to the emission of $\lambda 2.89 \mu$m 
photons has also been determined. With their percentage contributions
in parentheses, the most important terms are:  
$5 \; ^{3}\!D^{o}$ (29.4), $5 \; ^{3}\!S^{o}$ (13.7),
$6 \; ^{3}\!D^{o}$ (10.6), $6 \; ^{3}\!S^{o}$ (7.7),
$4 \; ^{3}\!D^{o}$ (6.9), and $7 \; ^{3}\!D^{o}$ (6.4)

\subsection{Predicted intensities}

With the quantities $w_{k}$ and $\epsilon^{k}_{ij}$ now calculated, 
surface brightnesses for all multiplets formed in the cascades
are obtained by summing equation (11) over the pumping transitions and over
the multiplets' components. The
results for all multiplets with $f_{\lambda} >\: $0.008 bu are given in
Tables 1 and 2. Table 1 lists the optical lines, with wavelengths in air
given in \AA.
Table 2 lists the infrared lines, with vacuum wavelengths given in $\mu$m.
The tabulated wavelengths $\lambda_{eff}$ are the multiplets' effective
wavelengths,
computed by weighting each component's $\lambda$ by that component's 
fractional contribution to $f_{\lambda}$.

	The cascades also result in a UV line spectrum extending almost
to the Lyman limit. These are not tabulated but we note that the two
strongest UV lines are the  $3s\; ^{3}\!S^{o} - 2p^{4}\; ^{3}\!P$ multiplet
at $\lambda$1304.28\AA\ and the $4d\; ^{3}\!D^{o} - 2p^{4}\; ^{3}\!P$
multiplet at $\lambda$973.02\AA. The predicted strengths of these lines are
16.7 and 2.11 bu, respectively.
 
	Of most immediate interest is the predicted strength 
for the  $4p\; ^{3}\!P - 4s\; ^{3}\!S^{o}$ O$\:${\sc i} multiplet
at $\lambda$2.89$\mu$m. From Table 2, we find that   
\begin{equation}
 f_{2.89\mu} = 0.14 \: bu      \;\;\; ,
\end{equation}
remarkably close to the maximum value of 0.12 bu measured by Rubin et al.
(2001) at position 1SW. Since the geometrical parameters $d$ = 0.2pc
and $\mu_{0}/\mu_{1}$ = 1 were chosen to represent this point, this outcome
is compelling evidence in favour of the proposed identification.

\section{Discussion}

	In this section, further testing of the proposed identification is
carried out using the theoretical understanding developed in Section 5.
In addition, the sensitivity of
of the results to various input parameters is briefly discussed.

\subsection{Theory-independent estimates}

	Because the prediction of about the right strength for the	
O$\:${\sc i} multiplet at $\lambda 2.89 \mu$m has required a rather
complicated, multi-parameter theory, it is of interest that this multiplet's
strength can be reliably estimated without invoking this theory. There are
two ways of doing this, both of which are independent of the line formation
mechanism and of the distribution of O$\:${\sc i} along the line-of-sight.

	The first estimate follows from noting that the proposed
identification has the same upper term ($4\; ^{3}\!P$) as the
O$\:${\sc i} line at $\lambda$4368\AA. Accordingly, whenever
the 4368\AA\
line is observed, we can confidently infer that emission at
$\lambda 2.89 \mu$m is also occurring and can compute the IR multiplet's
strength from
the formula $f_{2.89\mu} = f_{0.44\mu}$/1.48 - see Tables 1 and 2. This
calculation depends only on the A-values and frequencies of the transitions.

	For a definitive test of the O$\:${\sc i} identification using this
accurately known
intensity ratio, the measured optical and IR line strengths are needed at
exactly the same location. Unfortunately, this information seems not to be
available from
published spectra. Failing this, the approach adopted here is to use the
measured
strengths of the $\lambda$4368\AA\ line at locations near to 1SW to infer
strengths for the
$\lambda 2.89 \mu$m line. This can be done using
the data of Esteban et al. (1998).

 	 Esteban et al. (1998) report that the $\lambda$4368\AA\ multiplet
has reddening-corrected brightnesses of
0.036 and 0.060 bu at their slit positions 1 and 2. The inferred values
for the $\lambda 2.89 \mu$m line are therefore 0.024 and 0.040 bu, both
of which represent averages over the
\mbox{13.3 \arcsec $\times$ 2 \arcsec} slit and refer
to positions offset from $\theta^{1}$ Ori C by 45 and 27 \arcsec,
respectively.

	These inferred values are significantly less than the maximum of
0.12 bu 
observed by Rubin et al (2001) at 1SW, a position offset from 
$\theta^{1}$ Ori C  by 32 \arcsec.
But this latter measurement refers to a 0.74 arcsec$^{2}$ aperture extracted
from
their long-slit UKIRT spectrum. On the other hand, from their ISO SWS02
spectrum with a \mbox{14\arcsec $\times$ 20 \arcsec}
aperture also centered on 1SW, they report an average
brightness for the $\lambda 2.89 \mu$m line of 0.039 bu. This is in
satisfactory agreement with the spatial averages inferred from the
data of
Esteban et al. (1998) at similar offsets from the UV source. Accordingly,
these theory-independent estimates strongly support the identification of the
$\lambda 2.89\mu$m line as the $4p\; ^{3}\!P - 4s\; ^{3}\!S^{o}$ multiplet of
O$\:${\sc i}.

	The second theory-independent estimate derives from noting that other
observed optical O$\:${\sc i} lines tell us that terms higher than 
$4\; ^{3}\!P$ also emit. Accordingly, if the cascade from such a term results
in the emission of the $\lambda 2.89\mu$m multiplet, this emission's
intensity ratio to
the observed optical line can be computed;
and this allows a {\em contribution} to the  
strength of the $\lambda 2.89\mu$m line to be estimated from the strength
of the optical line. Note, however, that, when the excited upper term is
$^{3}\!D^{o}$, the intensity ratios depend on the relative populations of
the $^{3}\!D^{o}$ term's sublevels. Also the cascades overlap so the 
contributions to the $\lambda 2.89\mu$m line cannot simply be co-added.
Partly because of these complications, this means of estimating the strength
of the $\lambda 2.89\mu$m is not pursued further. But the main reason is that
it is more rewarding to use the information in the other optical  
O$\:${\sc i} lines to test the cascade theory of Section 5.

	These two theory-independent procedures can of course be applied
to infer the prescence of and to estimate stengths for numerous other
O$\:${\sc i}
lines in the IR. Since the observed optical lines at
$\lambda \lambda$5275 and 5299\AA\ imply excitation of levels as high as  
$7\;^{3}\!D^{o}$ and $8\;^{3}\!S^{o}$, most of the IR lines in Table 2 
must be present at some level at Positions 1 and 2 of Esteban et al. (1998).
But again, rather than proceed with such estimates, it is preferable to use
the data to test
the cascade theory. If the test is successful, {\em ratios} of line strengths
from Tables 1 and 2 should be reliable and so can be used to estimate
strengths of IR lines from observed optical lines.

\subsection{Cascade test}

As noted earlier, Grandi (1975) has previously demonstrated convincingly 
that pumped cascades account for the permitted O$\:${\sc i} triplet lines in  
optical spectra of the Orion Nebula. Neverthelees, it is of interest   
to test the cascade hypothesis again using the theoretical treatment given
here together with the modern CCD measurements of Esteban et al. (1998).

	The results of the cascade test are shown in Fig.5 where both the
observed and predicted line brightnesses are expressed as fractions of
the summed values for the lines $\lambda \lambda$4368, 5275, 5299, 5513,
5555, 5959 and 6046\AA. (The line $\lambda$7002\AA\ is excluded from this
sum because
it was observed only at Position 2.) Inspection of Fig. 5 reveals a
satisfactory degree of agreement 
between theory and observation for all but one of the eight observed lines.
The exception is the line at $\lambda$7002\AA.

\begin{figure}
\vspace{8.2cm}
\includegraphics{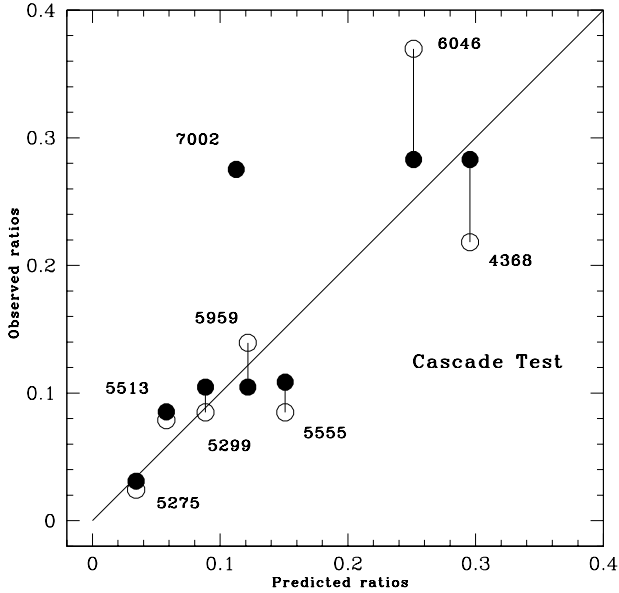}
\caption{Cascade test for permitted O$\:${\sc i} emission lines in the
optical spectrum of the Orion Nebula. Open and filled circles
refer to
Positions 1 and 2, respectively, of Esteban et al. (1998) and correspond to
the indicated wavelength in \AA. The theoretical
values are from Table 1.}

\end{figure}

	A point to note in assessing the success of this test is the
complexity of this cascade mechanism when compared to a simple recombination
cascade. In this case, predicted line strengths represent the superposition
of numerous separate pumped cascades whose relative importance is
determined by dust absorption through its effect on equivalent bandwidths
and escape probabilities. Evidently, there are interesting diagnostic
possibilities to be exploited using accurate measurements of the strengths
of these and other optical O$\:${\sc i} lines.  

	Note that none of the observed optical lines has a strength that
can be computed via a simple cascade calculation from one of the other 
optical lines. Thus Fig.5 is free from this species of degeneracy, which
further emphasizes its diagnostic importance.

\subsection{Wavelength}

To be accepted, a line identification must be consistent with the measured
wavelength of the uid line. From their ISO SWS02 spectrum, Rubin et al. 
(2001) infer a
rest wavelength of 2.89350 $\pm$ 0.00003$\mu$m, in essentially perfect
agreement with $\lambda_{eff}$ = 2.893509$\mu$m from Table 2.

	However, Rubin et al. derived this wavelength 
assuming that line formation occurs in the main H$\:${\sc ii} region
at $V_{helio}$ = 17 km s$^{-1}$. But for the proposed
identification, line formation occurs immediately behind the
ionization front, for which an appropriate velocity is 25.5 km s$^{-1}$
(O'Dell 2001a). Thus there is in fact a wavelength discrepancy equivalent to
8-9 km s$^{-1}$, significantly larger than the Rubin et al. error estimate of
3 km s$^{-1}$. But given that a theory-independent argument (Section 6.1)
places an O$\:${\sc i} line with about the right strength at this
slight velocity offset, this discrepancy is not grounds for
discarding the identification.

	A possible explanation of this velocity offset is that 
that the UV pumping occurs predominantly in neutral blobs entrained in the
outflowing ionized gas.

\subsection{Intensity profile}

Rubin et al. (2001) found that the highest surface brightness for the uid
line is coincident with the brightest part of the H$\:${\sc ii} region, and 
therefore understandably concluded that it originates from
within the main ionization zone. But the O$\:${\sc i} $\lambda 2.89 \mu$m
identification conflicts with this conclusion; and so there is the obvious
expectation that it will fail to yield a satisfactory distribution of
surface brightness.

	A crude calculation of the angular distribution of O$\:${\sc i}
cascade emission can be made using the 3-D model of the Orion Nebula 
constructed by Wen \& O'Dell (1995). Fig.5 of their paper gives the cross
section at position angle 226\degr, which includes the brightest part of the
Nebula. To a sufficient approximation, their profile for the ionization front
can be modelled as three straight line segments. Taking 500 pc as the
distance of the Nebula, we can use this model to compute the distance d in pc 
from $\theta^{1}$ Ori C to a point on the ionization front. The direction 
cosines $\mu_{0}$ and $\mu_{1}$ - see Fig.1 - at this point can also be
computed if we add the assumption that the ionization front is not tilted in
the direction perpendicular to this cross section. In this way,
the geometrical factor  
\begin{equation}
 C = \frac{\mu_{0}}{\mu_{1}} (\frac{0.2}{d})^{2}      \;\;\; ,
\end{equation}
can be derived and used 
to obtain the intensity profiles of the O$\:${\sc i}
cascade lines from the reference line brightnesses given in Tables 1 and 2.

	In Fig.6, the resulting intensity profile for the O$\:${\sc i}  
$\lambda 2.89 \mu$m multiplet is plotted for position
angle 226\degr. This
should be compared to Fig. 5a in Wen \& O'Dell (1995), which plots intensity
profiles for [N$\:${\sc ii}], H$\alpha$ and [O$\:${\sc iii}] lines at
this same
position angle. This comparison shows that O$\:${\sc i} line has a predicted
intensity profile that is qualitatively similar to those of lines emitted
by the H$\:${\sc ii} region. In particular, peak intensities at offsets of
$ \simeq 30 \arcsec$ to the SW are closely coincident, and all profiles show
a similar steep decline beyond their peaks in the SW. Thus the observed
intensity profile of the uid line, though suggestive of emission from within
the H$\:${\sc ii} region,
is in fact also consistent with line
formation by pumped cascades in the neutral zone immediately behind the
ionization front.

\begin{figure}
\vspace{4.7cm}
\includegraphics{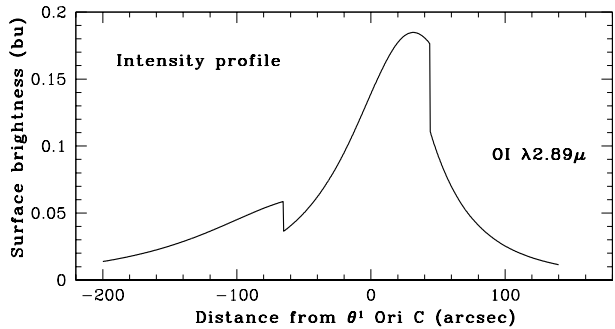}
\caption{Intesity profile for O$\:${\sc i} $\lambda 2.89 \mu$m emission.
This 1-D profile is at position angle 226\degr and corresponds to Fig. 5
in Wen \& O'Dell (1995).}

\end{figure}

	Note that the discontinuities in Fig.6 at offsets of -65 and +45
\arcsec are due to the discontinuities of $\mu_{0}/\mu_{1}$ at points where
the linear segments meet.

\subsection{Sensitivity}

In this subsection, the sensitivity of the cascade calculation to input
parameters is investigated. Given the strong evidence already presented in
support of
the O$\:${\sc i} identification, the main interest now is the
possible relevance of this line and, more importantly, of the optical
lines for diagnostic investigations. 

\subsubsection{Oxygen abundance}

An increase of $n(O)/n(H)$ by 0.1dex increases the strength of the 
$\lambda 2.89 \mu$m multiplet by 0.032dex, thus confirming the insensitivity
to abundance anticipated by the crude theory of Section 4.

	With respect to the observed optical lines, the responses to this
change range from 0.013dex ($\lambda$5959\AA) to 0.067dex ($\lambda$7002\AA).
This illustrates a useful differential sensitivity that can perhaps be
exploited diagnostically.

\subsubsection{Gas-to-dust ratio}

The number density of grains in the O$\:${\sc i} slab is determined by the
coefficient $\gamma$ in the empirical formula of Bohlin et al. (1978) -
see Section 5.2.1. Decreasing this number density by 1.0dex by increasing
$\gamma$ to 5.8 $\times$ 10$^{22}$ atoms cm$^{-2}$ mag$^{-1}$
results in a 0.23dex increase in the $\lambda 2.89 \mu$m line. The
corresponding changes in the optical lines range from 0.10dex 
($\lambda$5959\AA) to 0.46dex ($\lambda$7002\AA).

\subsubsection{Ratio of visual to selective extinction}

The dust absorption coefficient calculated in Section 5.2.1 assumes
$R_{V}$ = 3.1. This ratio is known to depend on
environment, with $R_{V}$ = 5
found in some dense clouds (e.g., Cardelli et al. 1989). With this value,
the strength of the $\lambda 2.89 \mu$m line increases by 0.044dex.
The corresponding changes in the optical lines range from 0.018dex 
($\lambda$5959\AA) to 0.091dex ($\lambda$7002\AA).

\subsubsection{Microturbulence}

The adopted kinematic model with $v_{D}$= 12.7 km s$^{-1}$ has been
described as {\em macro}turbulence because it does {\em not} apply on the
shortest length scales, namely the mean free paths of photons emitted in
decays to the O$\:${\sc i} ground term. On these scales, thermal broadening
has been assumed. If this assumption is dropped, and the Doppler parameter
in Section 5.3 also set = 12.7 km s$^{-1}$, then the kinematic model
becomes one of {\em micro}turbulence.

	With this change, 
the strength of the $\lambda 2.89 \mu$m line changes by -0.058dex.
The corresponding changes in the optical lines range from +0.001dex 
($\lambda$5959\AA) to -0.30dex ($\lambda$7002\AA). Notice that this change
would seriously worsen the already substantial discrepacy for the 
$\lambda$7002\AA\ line in Fig.5.

\subsubsection{Temperature of O$\:${\sc i} slab}

In the calculation of escape probabilities (Section 5.3), thermal broadening
is assumed at $T$ = 10000K. If this temperature is reduced to 3000K,
the $\lambda 2.89 \mu$m line's strength increases by 0.035dex.
The corresponding changes in the optical lines range from 0.005dex 
($\lambda$5959\AA) to 0.14dex ($\lambda$7002\AA).

\section{Conclusion}

The aim of this investigation has been to identify the strong
$\lambda 2.89 \mu$m emission line discovered by Rubin et al. (2001).
Surprisingly, the identification turns out to be a line from a neutral atom
(O$\:${\sc i}) rather than from an ion with a production ionization
potential between
13.6 and 54.4 eV as expected by Rubin et al. Correspondingly, the line
originates not from within the H$\:${\sc ii} region but from the neutral zone
behind the ionization front.
Because the Orion Nebula is a {\em thin} blister of ionized gas on the 
near side of the Orion Molecular Cloud (eg, O'Dell 2001b),  
lines-of-sight through the densest part of the H$\:${\sc ii} region 
also intersect the ionization front
and therefore also the neutral zone where UV pumping of O$\:${\sc i} occurs.
This explains why an O$\:${\sc i} cascade line can have a spatial intensity 
profile that 
mimics those of emission lines originating in the H$\:${\sc ii} region
(Section 6.4). 
 
	On a memorable previous occasion, puzzling uid nebular emission
lines  
also proved to be due to Oxygen (Bowen 1927), with momentous
consequences for astronomical spectroscopy (eg, Osterbrock 1988).
In the present case, by contrast,
one could argue that the identification is almost totally inconsequential
given
that that the strength of the $\lambda 2.89 \mu$m line can be inferred from
that of the $\lambda$4368\AA\ line (Section 6.1). However, the treatment of
O$\:${\sc i} cascades stimulated by this uid line reveals interesting
diagnostic possibilities, most or even all of which can be exploited with
measurements of the optical O$\:${\sc i} lines. First, the cascades occur
within a very
narrow zone immediately behind the ionization front. For example, with
$n(H) = 10^{5}$cm$^{-3}$ and $v_{D} = 12.7$km$\:$s$^{-1}$, the length scale
corresponding to $\tau = 1$ at the line
centre of the third strongest line pumping the $5 \; ^{3}\!D^{o}$ term      
is 4 $\times$ 10$^{-5}$pc, equivalent to an angular scale of about
0.02 \arcsec. Thus velocity measurements of the permitted O$\:${\sc i}
lines potentially give a sharply defined reference velocity for studies of
the dynamics of the Orion Nebula. There is also the possibility
of investigating the clumpiness of O$\:${\sc i} and therefore H$\:${\sc i}
at the point of photoionization by $\theta^{1}$ Ori C (Section 6.3).
In view of the abundance of Oxygen and the high $gf$-values of the pumping
transitions, this superb resolution in depth at ionization fronts may
never be surpassed.

	The second point concerning diagnostic potential is that intensity
ratios of optical lines are {\em not} simply derivable from A-values and
frequencies (Section 6.2): the astrophysics determining the relative
importance of the numerous independent cascades needs also to be correct;
and this
leads to these lines having usefully different sensitivities to input
parameters (Section 6.5).

	Yet another possibility is to exploit the geometrical factor $C$
- equation (24) - to determine the topography of the Orion
Nebula's ionization front independently of the procedure followed by Wen \&
O'Dell (1995). However, in such an investigation, it may be preferable
to replace the orientation factor $\mu_{0}/\mu_{1}$ by an average derived
from a statistical model of clumpiness.

	These intriguing diagnostic possibilities have not been
pursued in this paper
beyond noting that a substantial evacuation of grains from the cascade
zone reduces the $\lambda$7002\AA\ line's discrepacy in Fig.5, and that
a {\em micro}turbulent kinematic model would increase this discrepancy.

\section{Acknowledgements}

I am indebted to M.J.Barlow for bringing this problem to my attention. 
K.Butler, K.A.Berrington, S.N.Nahar and R.H.Rubin kindly provided
information and unpublished data. The final form of this paper benefited from
stimulating discussions with M.J.Barlow, X.-W.Liu and other members of the
nebular astrophysics group at UCL as well as from comments by the referee,
C.R.O'Dell.

\begin{table}
\caption{Optical O$\:${\sc i} lines formed in pumped cascades}

\begin{tabular}{ccc}
     $ \lambda_{eff}$          &  Multiplet  &  $f_{\lambda}$ (bu)   \\
                               &             &                      \\
     3433.43  &  $\:6p\; ^{3}\!P   - 3s\; ^{3}\!S^{o}$  &   0.0083  \\
     3692.38  &  $\:5p\; ^{3}\!P - 3s\; ^{3}\!S^{o}$  &   0.0427  \\
     4368.24  &  $\:4p\; ^{3}\!P - 3s\; ^{3}\!S^{o}$  &   0.2074  \\
     4925.65  &  $11d\; ^{3}\!D^{o} - 3p\; ^{3}\!P$ &   0.0080  \\
     4972.47  &  $10d\; ^{3}\!D^{o} - 3p\; ^{3}\!P$ &   0.0098  \\
     4980.05  &  $11s\; ^{3}\!S^{o} - 3p\; ^{3}\!P$ &   0.0175  \\
     5037.46  &  $\:9d\; ^{3}\!D^{o} - 3p\; ^{3}\!P$  &   0.0125  \\
     5047.74  &  $10s\; ^{3}\!S^{o} - 3p\; ^{3}\!P$ &   0.0259  \\
     5131.24  &  $8d\; ^{3}\!D^{o} - 3p\; ^{3}\!P$  &   0.0166  \\
     5146.57  &  $9s\; ^{3}\!S^{o} - 3p\; ^{3}\!P$  &   0.0391  \\
     5275.07  &  $7d\; ^{3}\!D^{o} - 3p\; ^{3}\!P$  &   0.0240  \\
     5299.00  &  $8s\; ^{3}\!S^{o} - 3p\; ^{3}\!P$  &   0.0621  \\
     5512.71  &  $6d\; ^{3}\!D^{o} - 3p\; ^{3}\!P$  &   0.0406  \\
     5554.95  &  $7s\; ^{3}\!S^{o} - 3p\; ^{3}\!P$  &   0.1059  \\
     5958.52  &  $5d\; ^{3}\!D^{o} - 3p\; ^{3}\!P$  &   0.0854  \\
     6046.38  &  $6s\; ^{3}\!S^{o} - 3p\; ^{3}\!P$  &   0.1764  \\
     7002.12  &  $4d\; ^{3}\!D^{o} - 3p\; ^{3}\!P$  &   0.0790  \\
     7254.36  &  $5s\; ^{3}\!S^{o} - 3p\; ^{3}\!P$  &   0.3007  \\
     8446.48  &  $3p\; ^{3}\!P - 3s\; ^{3}\!S^{o}$  &   2.1353  \\
     9723.43  &  $8p\; ^{3}\!P - 3d\; ^{3}\!D^{o}$  &   0.0094  \\

\end{tabular}
\end{table}

\begin{table}
\caption{Infrared O$\:${\sc i} lines formed in pumped cascades}

\begin{tabular}{ccc}
     $ \lambda_{eff}$          &  Multiplet  &  $f_{\lambda}$ (bu)   \\
                               &             &                      \\
        1.032313    &  $6p\; ^{3}\!P - 4s\; ^{3}\!S^{o}$  &  0.0142   \\    
        1.044628    &  $7p\; ^{3}\!P - 3d\; ^{3}\!D^{o}$  &  0.0131   \\    
        1.128987    &  $3d\; ^{3}\!D^{o} - 3p\; ^{3}\!P$  &  0.4272   \\    
        1.185827    &  $8d\; ^{3}\!D^{o} - 4p\; ^{3}\!P$  &  0.0103   \\    
        1.187145    &  $6p\; ^{3}\!P - 3d\; ^{3}\!D^{o}$  &  0.0173   \\    
        1.194037    &  $9s\; ^{3}\!S^{o} - 4p\; ^{3}\!P$  &  0.0083   \\    
        1.265549    &  $7d\; ^{3}\!D^{o} - 4p\; ^{3}\!P$  &  0.0148   \\    
        1.308046    &  $5p\; ^{3}\!P - 4s\; ^{3}\!S^{o}$  &  0.0378   \\    
        1.316817    &  $4s\; ^{3}\!S^{o} - 3p\; ^{3}\!P$  &  0.5303   \\    
        1.411486    &  $6d\; ^{3}\!D^{o} - 4p\; ^{3}\!P$  &  0.0270   \\    
        1.567012    &  $5p\; ^{3}\!P - 3d\; ^{3}\!D^{o}$  &  0.0202   \\    
        1.745830    &  $5d\; ^{3}\!D^{o} - 4p\; ^{3}\!P$  &  0.1028   \\    
        1.823426    &  $6s\; ^{3}\!S^{o} - 4p\; ^{3}\!P$  &  0.0320   \\    
        2.857113    &  $6p\; ^{3}\!P - 5s\; ^{3}\!S^{o}$  &  0.0083   \\    
        2.893509    &  $4p\; ^{3}\!P - 4s\; ^{3}\!S^{o}$  &  0.1401   \\    
        3.098498    &  $4d\; ^{3}\!D^{o} - 4p\; ^{3}\!P$  &  0.0133   \\    
        3.453334    &  $6d\; ^{3}\!D^{o} - 5p\; ^{3}\!P$  &  0.0302   \\    
        3.661737    &  $5s\; ^{3}\!S^{o} - 4p\; ^{3}\!P$  &  0.0430   \\    
        4.560809    &  $4p\; ^{3}\!P - 3d\; ^{3}\!D^{o}$  &  0.0216   \\    
        5.985350    &  $7d\; ^{3}\!D^{o} - 6p\; ^{3}\!P$   &  0.0102   \\    
        6.858507    &  $5p\; ^{3}\!P - 5s\; ^{3}\!S^{o}$   &  0.0238   \\

\end{tabular}
\end{table}

\end{document}